\documentclass[12pt]{article}
\usepackage{a4,latexsym,pslatex,graphicx,times,color,amsmath}
\usepackage{subfig} 

\usepackage[numbers]{natbib}

\title{An analytic velocity profile for pressure-driven flow of a Bingham fluid in a curved channel}
\author{T.G. Roberts and S.J. Cox \\ 
Department of Mathematics, Aberystwyth University, SY23 3BZ, UK.}
\date{December 2019}

\begin{document}

\maketitle

\begin{abstract}
We derive an expression for the velocity profile of a pressure-driven yield-stress fluid flowing around a two-dimensional concentric annulus. 
This result allows the prediction of the effects of channel curvature on the pressure gradient required to initiate flow for given yield stress, and for the width of the plug region and the flux through the channel at different curvatures.
We use it to validate numerical simulations of the flow from a straight channel into a curved channel which show how the fluid first yields everywhere before reaching the predicted velocity profile.
\end{abstract}

\section{Introduction}

Yield stress fluids are found in many situations, from toothpaste to drilling muds \cite{mousse13,barnes1989}. The property of the yield stress is often important, for example in preventing fluid flow in the absence of applied forces, and on the other hand often complicates applications, for example in requiring large stresses to be applied before a contaminated sludge can be processed. In enhanced oil recovery \cite{farajzadeh2012} the yield stress of a foam allows it to act as a displacement fluid, pushing oil in front of it. In such an application it is necessary to predict in which parts of the fluid the stress will exceed the yield stress, and the material will flow, and where the stress is so low that either the material does not move at all, or moves as a solid plug. 

A similar application is foam sclerotherapy \cite{nastasa2015}, a minimally-invasive treatment for varicose veins. Varicose veins are not only unsightly but also painful, and can often lead to further medical complications. Instead of, for example, surgery, such veins can be removed by injecting a sclerosant-laden foam into the affected vein, in much the same way as in enhanced oil recovery. The sclerosant must be delivered to the vein wall, without mixing with the blood that is present, to cause the vein to collapse and at the same time the foam must push the blood out of the vein. A foam with a high yield stress is therefore required for the treatment: a large plug region is required for effective blood displacement and to prevent too much mixing in the yielded regions close to the vein walls.

Perhaps the simplest example of a continuum model for a yield stress fluid is due to Bingham \cite{bingham1922},  almost a century ago. This model assumes zero strain rate below a critical value of the stress, and is therefore inelastic; this is a visco-plastic model.
This model has been extensively studied theoretically, for example for steady pressure-driven flow in straight channels of different cross-sections  \cite{birddy83,taylorw97,norouzivdbs15} and for boundary-driven flow in annuli~\cite{birddy83,muravlevemm10} and numerically, for example for flow past a sphere~\cite{blackerym97}.

A great deal of work on Bingham fluids is concerned with Couette flow~\cite{birddy83}, as in a Couette viscometer, in which the fluid is held between concentric cylinders and one of the cylinders moves tangentially. Away from the laboratory, many flows of yield stress fluids are pressure-driven, often in curved or bent pipes~\cite{speddingbm04}. To the best of our knowledge, closed form analytic solutions for pressure-driven flow in an annulus have not been previously derived. 
Norouzi et al. \cite{norouzivdbs15} proposed the use of an infinite series solution for the velocity profile in a curved three-dimensional channel with a rectangular cross-section. We take a different approach and, for simplicity, consider the equivalent 2D case, but seek a closed form expression for the velocity and stress profiles.

We consider the slow 2D pressure-driven (Poiseuille) flow of a Bingham fluid in a curved duct. The Dean number is assumed to be small (``creeping'' Dean flow), so that we neglect inertia, centripetal forces and any consequent secondary flows. In \S \ref{sec:maths} we give the governing equations of the flow, the constitutive equation for the fluid, and outline our solution, which requires determination of the radial positions  of the yield surfaces. We describe predictions for the velocity and stress fields in \S \ref{sec:results}. In \S \ref{sec:simdat} we describe numerical simulations of the flow from a straight channel into an annulus, which describe the distance over which the velocity profile makes the transition from one solution to another. Finally, in \S \ref{sec:concs} we discuss the implications of our work for flow in narrow curved channels, such as occurs during the process of varicose vein sclerotherapy.

\section{Mathematical model}
\label{sec:maths}

\begin{figure}
\centerline{
\includegraphics[width=0.5\textwidth]{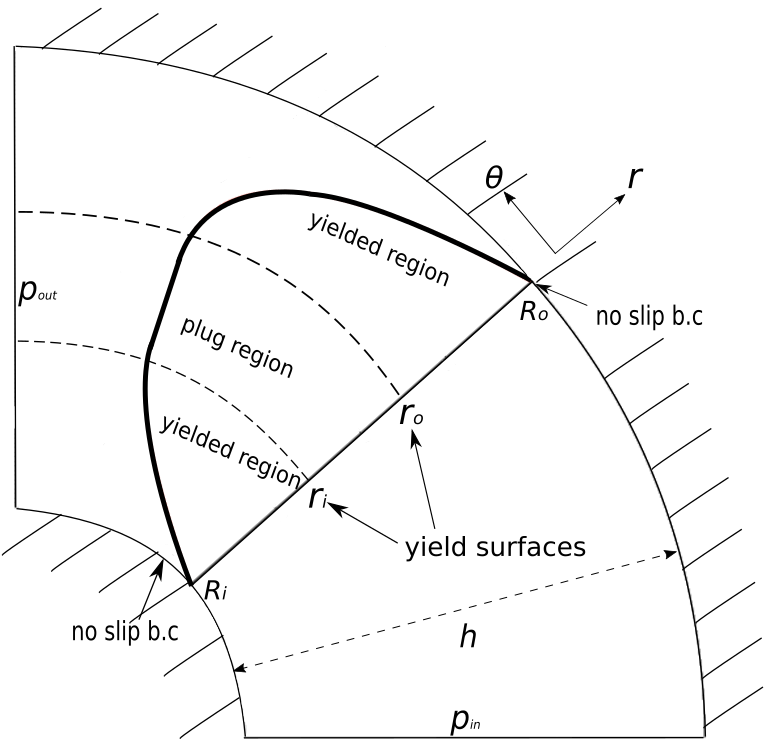}
}
\caption{The diagram indicates the geometry of the channel under consideration. Relative to plane polar coordinates $r$ and $\theta$, the channel has inner radius $R_i$ and outer radius $R_o$. Fluid flows in the positive $\theta$ direction due to a pressure difference $p_{\rm in} - p_{\rm out}$. An example of a velocity profile is shown in red, with a plug region between yield surfaces at $r=r_i$ and $r=r_o$.}
\label{fig:setup}
\end{figure}

\subsection{Governing equations}

We consider the steady, unidirectional flow of a Bingham fluid in the annular channel shown in Fig.~\ref{fig:setup}, described by polar coordinates $r$ and $\theta$. The annulus has inner and outer radii  $R_i$ and $R_o$ respectively, giving a channel of width  $h=R_o-R_i$. 

The fluid moves in response to a constant pressure gradient $G$ acting in the $\theta$-direction, which can be written in terms of the inlet and outlet pressures $p_{in}$ and $p_{out}$ and the position of the centreline of the channel, $R_{c} = \frac{1}{2} \left(R_i + R_o\right)$ as $G = (p_{out}-p_{in})/\theta R_c$. Note therefore that the pressure gradient $G$ always appears with the length-scale $R_c$ \cite{norouzivdbs15, rieger1973creeping} to take into account that the distance between the ends of the annular region increases with $r$, and therefore that the pressure gradient should decrease with increasing radial position. 

According to Stokes' equations this pressure gradient is balanced by the divergence of the stress. For this unidirectional flow the only non-zero component of the stress tensor $\underline{\underline{\tau}}$ is $\tau_{r \theta}$, so this becomes
\begin{equation} \label{eq:stokes_dim}
- \frac{R_c}{r} G =  \frac{1}{{r}^2} \frac{\partial}{\partial r} \left( {r}^2 \tau_{r \theta} \right).
\end{equation}
Similarly, the only non-zero component of the rate-of-strain tensor $\dot{\underline{\underline{\gamma}}}$ is the $r \theta$ component and the fluid velocity is $u_{\theta}(r)$. In consequence, the tensorial form of the constitutive equation for a Bingham fluid~\cite{denn2011issues} simplifies and the condition for yielding no longer requires calculation of the second invariant of the rate-of-strain tensor but becomes simply $|\tau_{r \theta}| > \tau_y $. The constitutive equation is therefore linear:
\begin{equation} 
\label{eq:constitutivedimensional}
\begin{array}{ll}
\tau_{r \theta}  = {\color{red} \pm}\tau_y + \mu r \displaystyle \frac{\partial}{\partial r}  \left(\displaystyle \frac{u_{\theta}}{r} \right)  & \quad \text{for} \quad |\tau_{r \theta}| > \tau_y \\
\dot{\gamma}_{r \theta} = 0 & \quad \text{for} \quad |\tau_{r \theta}| \leq \tau_y.
\end{array}
\end{equation}
We then consider three distinct regions in the flow. The sign in front of the Bingham number is positive in the inner yielded region, where the stress is positive, and negative in the outer yielded region, where the stress is negative. In the centre of the channel, where the magnitude of the shear stress $\tau_{r\theta}$ is below the yield stress, there is a ``plug" of fluid with zero strain-rate. At the wall we impose a no-slip boundary condition, $u_\theta(R_i) = u_\theta(R_o) = 0$, so that at each side of the plug region, close to the walls of the channel, the magnitude of the shear stress is greater and the fluid yields. In these two regions the velocity profile is parabolic, while in the plug region the fluid undergoes solid body rotation with $u_{\theta}$ proportional to $r$.

We consider the governing equations in dimensionless form relative to the length-scale $h$ and the velocity scale $U = G h^2 / \mu$.  Denoting dimensionless quantities with an asterisk we use
\begin{equation}
r^* = \frac{r}{h}, \quad \boldsymbol{u}^*_{\theta} = \frac{\boldsymbol{u}_{\theta} \mu}{G h^2}, \quad \nabla^* = h \nabla,  \quad \boldsymbol{\tau}^* = \frac{\boldsymbol{\tau}}{G h}.
\end{equation}
Introducing the channel curvature $\kappa = h / R_i$, the momentum balance eq.~(\ref{eq:stokes_dim}) becomes
\begin{equation} \label{eq:stokes}
-\left(\frac{1}{\kappa} + \frac{1}{2} \right) =  \frac{1}{{r^*}} \frac{\partial}{\partial r^*} \left( {r^*}^2 \tau_{r \theta}^* \right),
\end{equation}
where the left hand side arises from writing $R_c = R_i + h/2$.
The constitutive equation (\ref{eq:constitutivedimensional}) becomes
\begin{equation} 
\label{eq:constitutive}
\begin{array}{ll}
 \tau_{r \theta}^*  =  \pm \frac{1}{2}B + r^* \displaystyle \frac{\partial}{\partial r^*}  \left(\displaystyle \frac{u_{\theta}^*}{r^*} \right)  & \quad \text{for} \quad |\tau_{r \theta}^*| > B/2 \\
\dot{\gamma}_{r \theta}^* = 0 & \quad \text{for} \quad |\tau_{r \theta}^*| \leq B/2.
\end{array}
\end{equation}
The yield-stress $\tau_y$ and viscosity $\mu$ are absorbed into a dimensionless Bingham number, representing the ratio of the yield stress to the viscous stresses \cite{bingham1922}:
\begin{equation} \label{eq:bing}
B = \frac{2\tau_y}{G h}.
\end{equation}
For small values of $B$ the profiles of velocity and stress will be similar to those for a Newtonian fluid of comparable viscosity. Increasing $B$ at fixed pressure gradient causes a widening plug region to develop in the centre of the channel and results in a decrease in the fluid flux.

From this point onwards, we drop the asterisks denoting dimensionless quantities.

\subsection{Analytic solution}

In the yielded regions, the solution to eq. (\ref{eq:stokes}) takes the form
\begin{equation}
\tau_{r\theta} = -\left(\frac{ 2 + \kappa}{4 \kappa} \right)  +  \frac{C}{r^2},
\label{eq:stress1}
\end{equation}
where $C$ is a constant of integration. In principle the constant of integration could be different in each region of the flow, but matching the stresses at each yield surface, or by applying a balance between pressure and stress at a selected control volume~\cite{laird1957slurry}, indicates that they are equal. 
Therefore the shear stress decreases quadratically across the gap, taking its maximum value at the inner wall $r= 1/\kappa$,  where the pressure gradient is greatest.
In the absence of a yield stress, $B = 0$, the velocity profile for a Newtonian fluid in this geometry is

\begin{equation}
u_{\theta}(r) = \frac{1}{4} \left(\frac{1}{\kappa} + 1\right)^2 \ln \left(\frac{1}{\kappa} + 1 \right) \left( r - \frac{1}{\kappa^2 r} \right) - \frac{1}{4} \left(\frac{1}{\kappa}\right)^2 \ln \left(\frac{1}{\kappa} \right) \left( r - \left( \frac{1}{\kappa} + 1\right)^2 \frac{1}{r} \right) - \left(\frac{2 + \kappa}{4 \kappa} \right) r \ln \left( r \right)
.
\label{eq:velNewt}
\end{equation}
This profile provides a reference state for the more general case.

According to the constitutive equation, eq.~(\ref{eq:constitutive}), the fluid yields when the magnitude of the shear stress is equal to $B/2$. We can therefore find the positions of the inner and outer yield surfaces, $r_i$ and $r_o$, at the points where $\tau_{r\theta} = B/2$ and $-B/2$ respectively:
\begin{equation}
r_i^2 = \frac{2 C}{ \left(\frac{ 2 + \kappa}{2 \kappa} \right) + B} , \quad  r_o^2 = \frac{2C}{ \left(\frac{ 2 + \kappa}{2 \kappa} \right) - B}.
\end{equation}
Eliminating $C$ gives a relationship between the positions of the inner and outer yield surfaces, written in terms of modified Bingham numbers $B^{\pm} =  \frac{1}{2} \left(\left(\frac{ 2 + \kappa}{2 \kappa} \right) \pm B \right) $:
\begin{equation} \label{eq:yieldsurfrel}
B^{+} r_i^2 = B^{-} r_o^2 .
\end{equation}
In addition, substituting for $C$ in eq.~(\ref{eq:stress1}) gives two equivalent expressions for the stress in terms of the position of either yield surface:
\begin{equation} \label{eq:stress}
\tau_{r\theta} = - \left(\frac{ 2 + \kappa}{4 \kappa} \right) + B^+ \frac{{r_i}^2}{r^2}  \quad \mbox{and} \quad  
\tau_{r\theta} = -\left(\frac{ 2 + \kappa}{4 \kappa} \right) + B^- \frac{{r_o}^2}{r^2}.
\end{equation}
These only apply in each of the two yielded regions of the flow, $\frac{1}{\kappa} \le r \le r_i$ and $r_o \le r \le \frac{1}{\kappa} + 1$, where substitution into the constitutive equation (\ref{eq:constitutive}) gives the velocity profile there.

Between these regions the fluid moves in a solid-like plug.
In this region of zero strain-rate, $\frac{\partial}{\partial r} \left(\frac{u_{\theta}}{r} \right) = 0$, which implies $u_{\theta} = A r$ with $A$ a constant found by ensuring that the velocity is continuous at the yield surfaces.

Applying no-slip boundary conditions at the walls $r=\frac{1}{\kappa}$ and $r=\frac{1}{\kappa} + 1$
then gives the velocity profile itself:
\begin{equation} 
\label{eq:velprof}
u_{\theta}(r) = \left\{
\begin{array}{ll}
B^+ \left(
    \displaystyle \frac{r_i^2}{2} 
       \left(\displaystyle r \kappa^2 -\displaystyle\frac{1}{r} \right) 
        -r \ln\left(\displaystyle r \kappa \right) 
  \right) & 
   \text{for} \quad \frac{1}{\kappa} \leq r \le r_i \\
B^+ r \left(
    \displaystyle\frac{1}{2} \left(\displaystyle r_i^2 \kappa^2 - 1 \right)
     -\ln\left(\displaystyle r_i \kappa \right) 
     \right) & 
   \text{for}\quad   r_i \leq r \leq r_o \\
B^- \left( 
    \displaystyle\frac{r_o^2}{2}
      \left(\displaystyle\frac{r \kappa^2}{(1 + \kappa)^2} - \displaystyle\frac{1}{r} \right)
      - r \ln\left(\displaystyle\frac{r \kappa }{1 + \kappa}\right) 
      \right) &  
\text{for} \quad r_o \le r \leq \frac{1}{\kappa} + 1 ,
\end{array}
\right.
\end{equation}
as well as a condition to determine the position of the inner yield surface $r_i$:
\begin{equation} \label{eq:yieldsurf}
B^{+} \left(- \ln\left(r_i \kappa \right) + \frac{1}{2} \left(r_i^2 \kappa^2 - 1 \right) \right)  =  B^{-} \left( - \ln \left(\sqrt{\frac{B^{+}}{B^{-}}} \frac{{r_i \kappa}}{\kappa + 1} \right) + \frac{1}{2} \left( \frac{B^{+}}{B^{-}} \frac{{r_i}^2 \kappa^2}{(\kappa + 1)^2} - 1\right) \right).
\end{equation}
Having found $r_i$, eq. (\ref{eq:yieldsurfrel}) gives the position of the outer yield surface $r_o$. 

To solve eq.~(\ref{eq:yieldsurf}) for $r_i$ we collect terms in $\ln(r_i)$ and $r_i^2$ to write it in the form
\begin{equation} \label{eq:yieldsurfeqn}
-2\alpha \ln(r_i) + r_i^2 + \beta = 0.
\end{equation}
The constants are
\begin{equation} 
\alpha = \frac{2B (\kappa + 1)^2 }{ B^+ \kappa^3 (\kappa + 2)},
\end{equation}
\begin{equation} 
\beta = \frac{2 (\kappa + 1)^2}{ B^+ \kappa^3 \left( \kappa + 2 \right)} \left(-B - B^+ \ln(\kappa) + B^- \ln\left(\sqrt{\displaystyle\frac{B^+}{B^-}} \displaystyle \frac{\kappa}{\kappa + 1} \right) \right).
\end{equation}
Equations of the form (\ref{eq:yieldsurfeqn}) have an exact solution in terms of the Lambert $W$ function \cite{corless1996lambertw}:
\begin{equation} \label{eq:riexactsoln}
r_i = \sqrt{-\alpha \;\; W \left( -1, -\frac{1}{\alpha} \exp \left(\frac{\beta}{\alpha}\right) \right)},
\end{equation}
where the $-1$ branch is used because the second argument is negative.

Alternatively, it is straightforward to find the root $r_i$ numerically using a root-finding algorithm to solve eq. (\ref{eq:yieldsurf}).

This provides the necessary input to give the velocity profile $u_\theta(r) $ in equation~(\ref{eq:velprof}) in terms of the channel curvature $\kappa$ and the Bingham number $B$.

\subsection{Constraints on the solution}

A Bingham fluid flowing through a straight channel of width  $1$ due to a unit pressure gradient will flow provided that $B$ is less than $1$. That is, below a critical Bingham number $B_c$, the shear stress induced at the walls of the channel will not exceed the yield stress, and then the material will not move. Using $y$ for the cross-channel coordinate, the critical value is found when the yield surfaces at $y = \pm B/2$ reach the walls at $y = \pm 1/2$.

In the curved channel  that we consider here, an indication that flow can cease is that for the velocity profile to be defined in eq.~(\ref{eq:yieldsurf}) we must have $B^-$ positive, giving an upper bound for the Bingham number: 
\begin{equation} \label{eq:Bup}
 B \leq \frac{2 + \kappa}{2 \kappa},
\end{equation}
indicated by the dashed line in Fig. 2.

\begin{figure}
\centering
\includegraphics[width = 7.5cm]{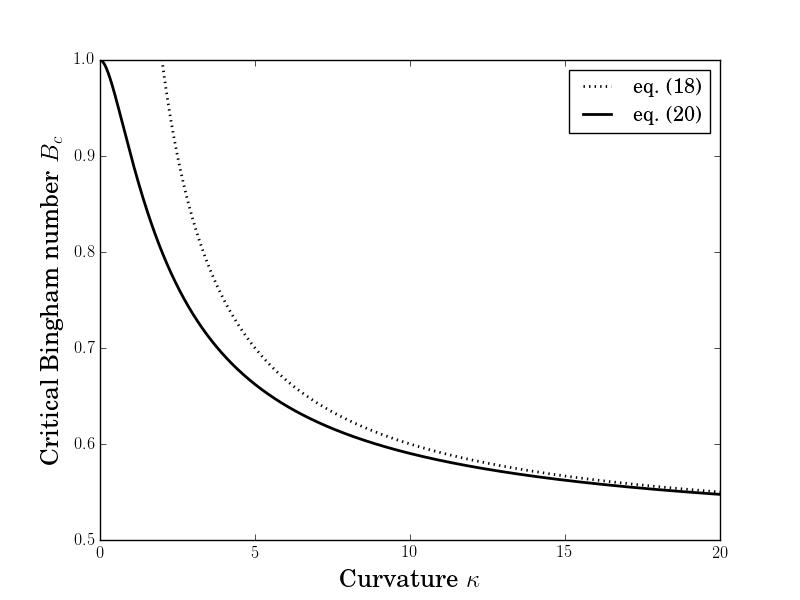}
\caption{The value of the critical Bingham number $B_c$ (eq. (\ref{eq:crit2B/Gh})), above which flow ceases, decreases with increasing channel curvature $\kappa$.  The dashed line indicates the upper bound for $B$ in eq. (\ref{eq:Bup}).}
\label{fig:critB}
\end{figure}

When the fluid is stationary everywhere the yield surfaces coincide with the walls of the channel, i.e. $r_i = \frac{1}{\kappa} $ and $r_o = \frac{1}{\kappa} + 1$.
Then eq.~(\ref{eq:yieldsurfrel}) gives
\begin{equation}
\left( B + \left( \frac{1}{\kappa} + \frac{1}{2} \right) \right) \left(\frac{1}{\kappa}\right)^2 = 
\left( -B + \left( \frac{1}{\kappa} + \frac{1}{2} \right) \right) \left( \frac{1}{\kappa} + 1 \right)^2.
\end{equation}
This can be rearranged to give a critical Bingham number
\begin{equation} \label{eq:crit2B/Gh}
B_c =  1 - \frac{\kappa^2}{2((\kappa + 1)^2 + 1)},
\end{equation}
shown in figure~\ref{fig:critB}, above which the flow stops.
For small $\kappa$, $B_c$ tends to one and we recover the result for the straight channel. 
As the channel curvature increases, either through reducing the inner radius $R_i$ or increasing the width $h$,  the value of $B_c$ is reduced towards a value of 0.5, indicating that a larger pressure gradient is required to induce flow. Thus a small amount of channel curvature has a surprisingly large effect on inhibiting flow. 
 
\section{Results}
\label{sec:results}

In presenting our results we scale the Bingham number $B$ by its critical value $B_c$, to represent the magnitude of the yield stress, and shift radial position to represent distance from the inner wall of the channel, $\hat{r} = r - \frac{1}{\kappa}$.

\subsection{Velocity}
\label{sec:results_velocity}

\begin{figure}
\centering
\subfloat[]{{\includegraphics[width=7.5cm]{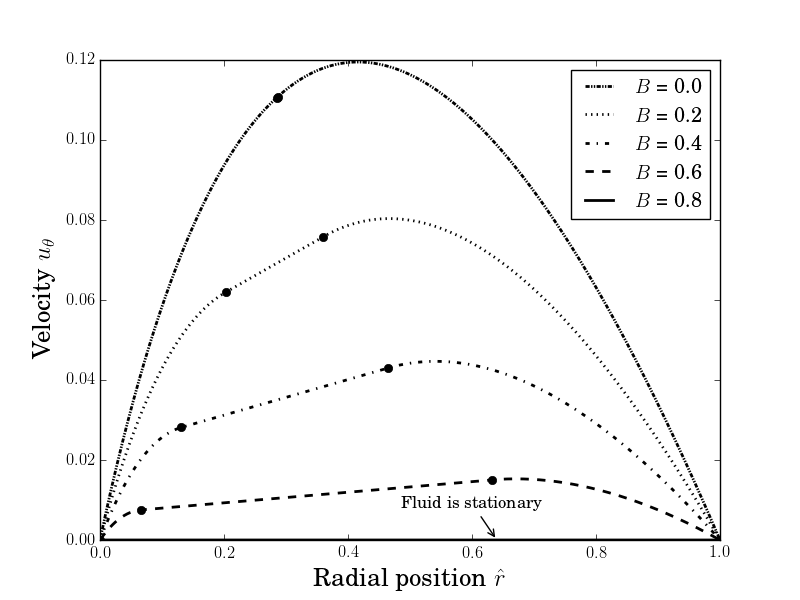} }}%
\qquad
\subfloat[]{{\includegraphics[width=7.5cm]{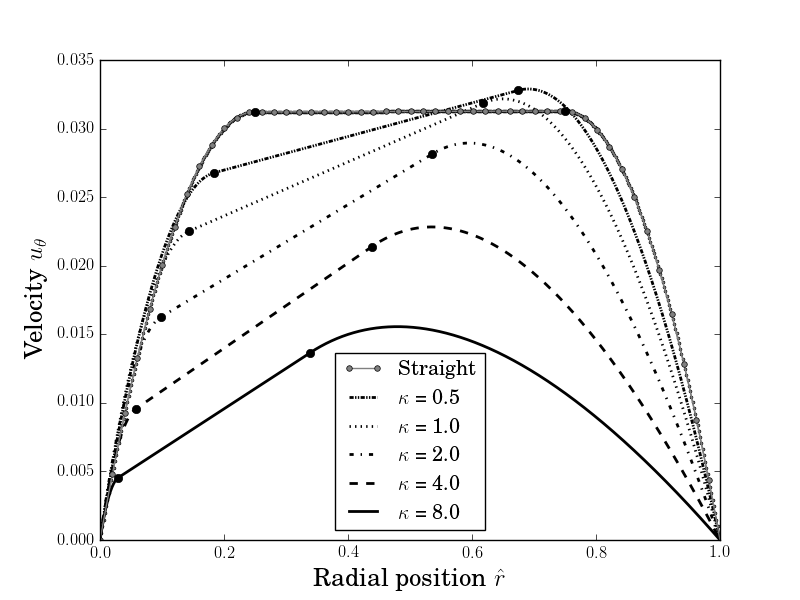} }}%
\caption{
Velocity profile in a curved channel as a function of radial position:
(a) for fixed channel curvature $\kappa = 2$ (for which $B_c = 0.8$) and different Bingham numbers;
(b) for constant Bingham number $B= 0.5$ for different channel curvatures $\kappa$.
The dots mark the positions of the yield surfaces.
}
\label{fig:velocity}%
\end{figure}

For a given Bingham number $B$ and channel geometry set by $\kappa$, we can find the  positions of both yield surfaces, which are discussed in \S \ref{sec:results_surface}, and hence plot the velocity  profile, eq. (\ref{eq:velprof}), which is shown in Fig. \ref{fig:velocity}.
The result (Fig. \ref{fig:velocity}(a)) is that the fluid velocity decreases and the plug width increases as the yield stress is increased. As $B$ approaches $B_c$, the flow stops.

Fig. \ref{fig:velocity}(b) shows how the velocity profile is affected by changes in channel curvature with fixed fluid properties $B$.  Increasing the channel curvature $\kappa$ reduces the velocity, particularly in the inner half of the channel. 
The slope of the velocity in the plug region is high for large curvatures $\kappa$ corresponding to high curvature of the stress profile. In the limit $\kappa \rightarrow 0$ we obtain the velocity profile for the straight channel case~\cite{birddy83} with position $\hat{y} \in [0,1]$:
\begin{equation} \label{eq:sc_vel}
u_x (y) = \left\{
\begin{array}{ll}
\frac{1}{2} \hat{y} \left( 1 - \hat{y} \right) - \frac{1}{2} B \left(\frac{1}{2} - |\hat{y}-\frac{1}{2}| \right)
& \text{for} \quad |\hat{y}-\frac{1}{2}| \geq \frac{B}{2} \\
\frac{1}{8} \left( 1 - B^2 \right) - \frac{1}{4} B \left(1 - B \right) &  
\text{for} \quad |\hat{y}-\frac{1}{2}| < \frac{B}{2} ,
\end{array}
\right.
\end{equation}
which is included in Fig. \ref{fig:velocity}(b).

The point of maximum velocity, close to $r=r_o$, moves away from the outer wall as $\kappa$ increases, but the value of the velocity there does not change monotonically: for intermediate curvature (e.g. $\kappa = 1.0$) the maximum velocity of the fluid  exceeds the value  in a straight channel.

\begin{figure}
\centering
\subfloat[]{{\includegraphics[width=7.5cm]{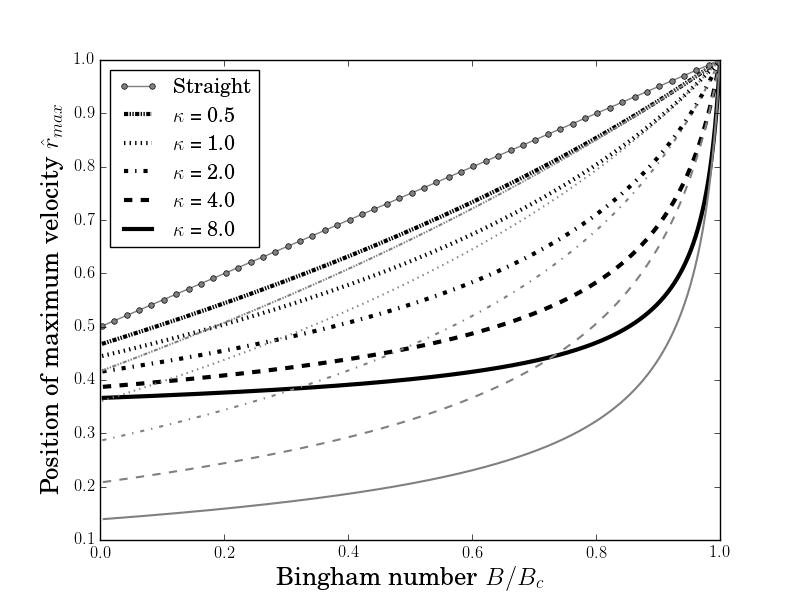} }}%
\qquad
\subfloat[]{{\includegraphics[width=7.5cm]{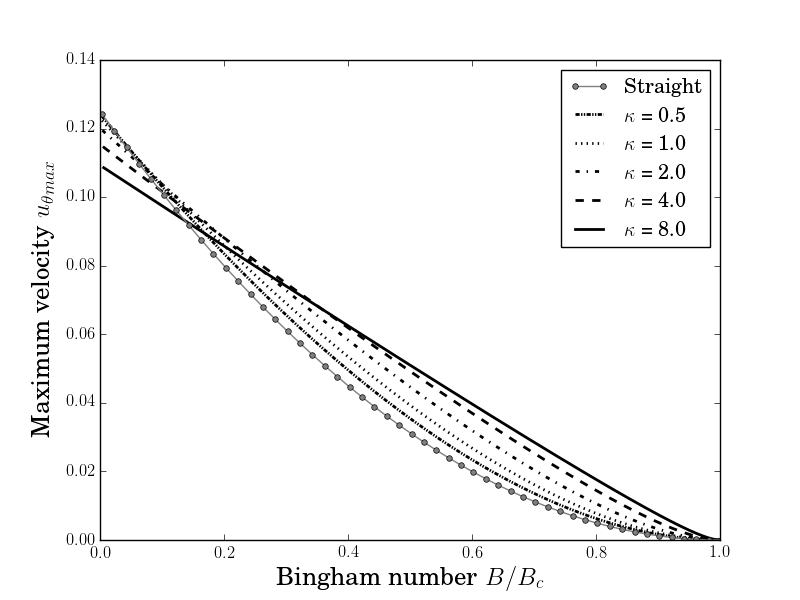} }}%
\caption{
The point of maximum velocity $r_{max}$ with fixed channel curvatures $\kappa$ as a function of Bingam number $B/B_c$. 
(a) The radial position of the maximum velocity (thick lines), compared with the position of the outer yield surface (thin lines) for several values of channel curvature $\kappa$.
(b) The value of the maximum velocity when $r = r_{max}$.
 }%
\label{fig:maxvelpos}%
\end{figure}

Figure~\ref{fig:maxvelpos}(a) shows that the radial position of the point of maximum velocity is always greater than or equal to $r_o$. Differentiation of the velocity profile (eq.~\ref{eq:velprof}) appropriate to radial positions $r_o \leq r \leq \frac{1}{\kappa} + 1$ gives:
\begin{equation}
\frac{d}{dr}(u_\theta(r)) = B^{-} \left( \frac{r_o^2}{2}\left( \frac{\kappa^2}{(\kappa + 1)^2} + \frac{1}{r^2} \right) - \left( \ln\left(\frac{r \kappa}{\kappa + 1}\right) + 1 \right) \right).
\end{equation}
Equating this to zero gives an expression for the radial position of maximum velocity $r_{max}$:
\begin{equation} \label{eq:r_max_poly}
\ln\left(\frac{r_{max} \kappa}{\kappa + 1}\right) + 1 = \frac{r_o^2}{2} \left( \frac{1}{r_{max}^2} + \frac{\kappa^2}{ (\kappa + 1)^2} \right).
\end{equation}
Using the Lambert W-function in the same way as used to find the inner yield surface $r_i$ in eq. (\ref{eq:riexactsoln}), we can rearrange eq. (\ref{eq:r_max_poly}) into an equation of the form:
\begin{equation} \label{eq:r_max_eqn}
\gamma \ln\left(\frac{1}{r_{max}^{2}}\right) + \frac{1}{r_{max}^{2}} + \zeta = 0
\end{equation}
where constants $\gamma$ and $\zeta$ are
\begin{equation} 
\gamma = \frac{1}{r_o^2}, \quad \quad \zeta =  \frac{1}{(1/\kappa + 1)^2} + \frac{2}{r_o^2} \left( \ln(1 /\kappa + 1) -1\right).
\end{equation}
Then the radial position of the maximum velocity is
\begin{equation} \label{eq:r_max}
r_{max} = \frac{1}{\sqrt{\gamma \; W \left( 0, \frac{1}{\gamma} \exp(-\zeta / \gamma) \right) }},
\end{equation}
where the $0$ branch is used because the second argument is positive. 

The value of $r_{max}$ from eq. (\ref{eq:r_max}) is shown in Fig. \ref{fig:maxvelpos}(a).
As channel curvature $\kappa$ decreases, the point of maximum velocity approaches a linear interpolation between the midpoint of the channel for $B=0$ and the outside of the channel for $B=B_c$, as for a straight channel.
As $B$ increases relative to $B_c$, the slope of the velocity profile in the plug region is reduced (and the fluid velocity also decreases). Hence $r_{max}$ approaches the outer yield surface and they eventually coalesce (Fig. \ref{fig:maxvelpos}(a)). For less-curved channels, this coalescence is seen at smaller values of $B/B_c$.

The value of the maximum velocity itself, ${u_{\theta}}_{max}$, is shown in Fig. \ref{fig:maxvelpos}(b) for different channel curvatures $\kappa$ as the Bingham number $B$ changes.
For $B = 0$, i.e. a Newtonian fluid, the maximum velocity of the fluid increases as the channel curvature decreases (smaller $\kappa$). As $B/B_c$ is increased, we notice a crossover where channels with greater curvature  induce a flow with a higher maximum velocity, and this point occurs at a radial position further away from the outer wall. 


At the crossover, the yield-stress is relatively small, but nonetheless indicates the competition between the curvature of the channel and the yield stress of the fluid in determining the fluid motion. The position of the point of maximum velocity is far enough from the outer wall that the no slip condition is not dominant, but not so close to the inner wall that the higher curvature there induces larger stresses.

\subsection{Yield surface positions and plug width}
\label{sec:results_surface}

\begin{figure}
\centering
\includegraphics[width=7.5cm]{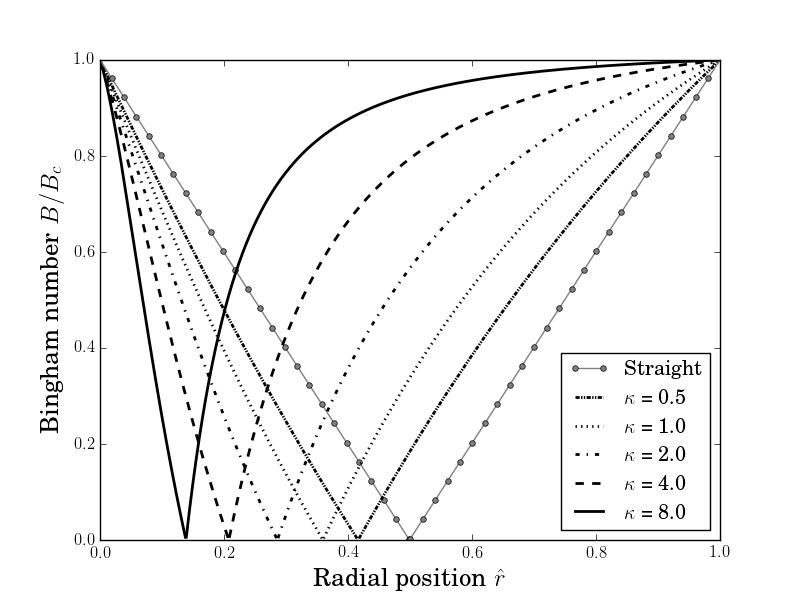}
\caption{Yield surface positions as a function of Bingham number $B/B_c$ for different values of channel curvature $\kappa$.
}%
\label{fig:ys}%
\end{figure}

The radial positions of the yield surfaces, from eq. \ref{eq:riexactsoln}, are shown in Fig.~\ref{fig:ys}. 
In the limit  $B \rightarrow 0$, the material behaves like a Newtonian fluid:  there are no yield surfaces and the values $r_i$ and $r_o$ coincide at a point close to the middle of the channel. 
As the curvature of the channel increases (Fig. \ref{fig:ys}) this point moves towards the inner wall. As $\kappa \rightarrow 0$ they meet at $\hat{r} = \frac{1}{\kappa} + \frac{1}{2}$, as in a straight channel.

As $B$ increases, the two yield surfaces move apart, reaching the inner and outer walls precisely when $B$ reaches $B_c$.  For large channel curvatures $\kappa$ the outer yield surface remains close to the centre of the channel until $B$ reaches about half of its critical value, while the position of the inner yield surface is almost linear in $B/B_c$ in all cases.

\begin{figure}
\centering
\includegraphics[width=7.5cm]{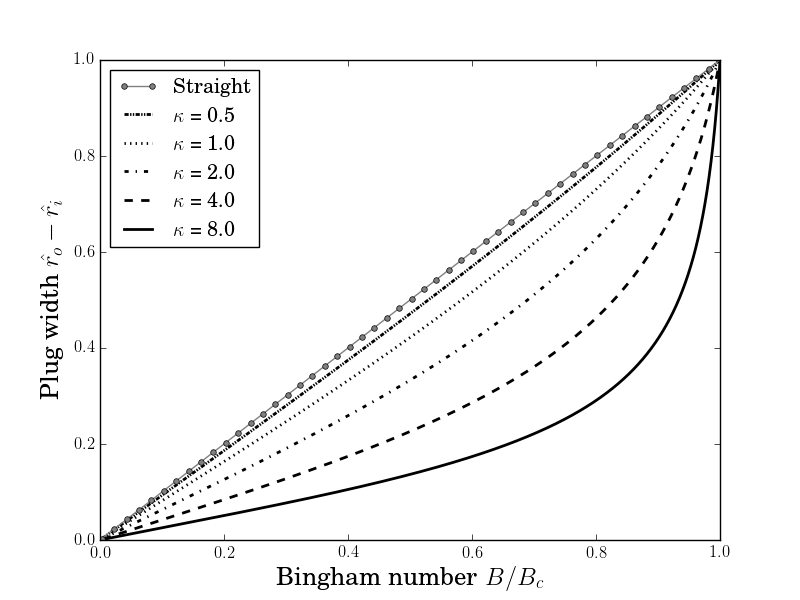}
\caption{Plug width as a function of Bingham number $B/B_c$ for different values of channel curvature $\kappa$.
}
\label{fig:pluglength}%
\end{figure}

The distance between the yield surfaces is the plug width, the region of low stress in which the material moves as a solid body, shown in Fig.~\ref{fig:pluglength}.  As  $B  \rightarrow B_c $ the plug width increases until it spans the whole channel.  For channels of larger curvature $\kappa$, the plug width increases more slowly with $B$ and then sharply increases as $B$ approaches $B_c$. 
For weakly curved channels (small $\kappa $), the plug width becomes linear in $B/B_c$.

\subsection{Shear stress}

\begin{figure}
\centering
\subfloat[]{{\includegraphics[width=7.5cm]{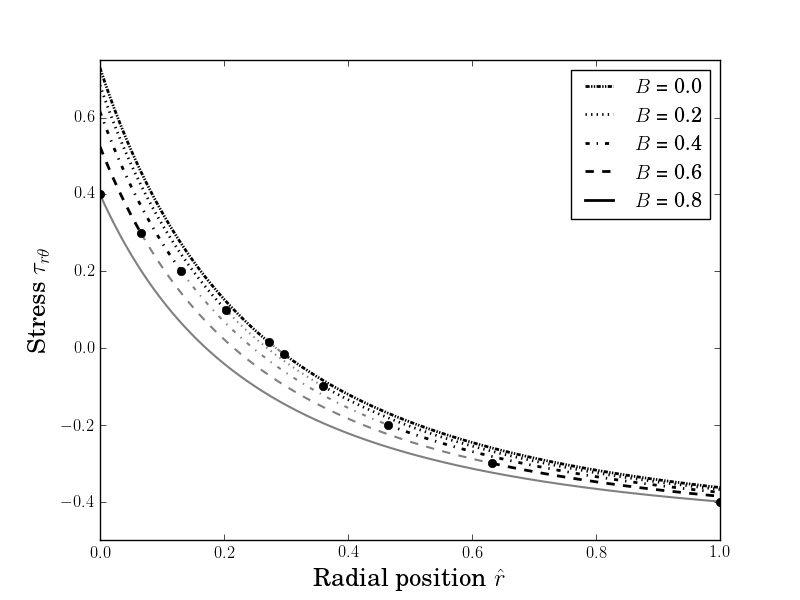} }}%
\qquad
\subfloat[]{{\includegraphics[width=7.5cm]{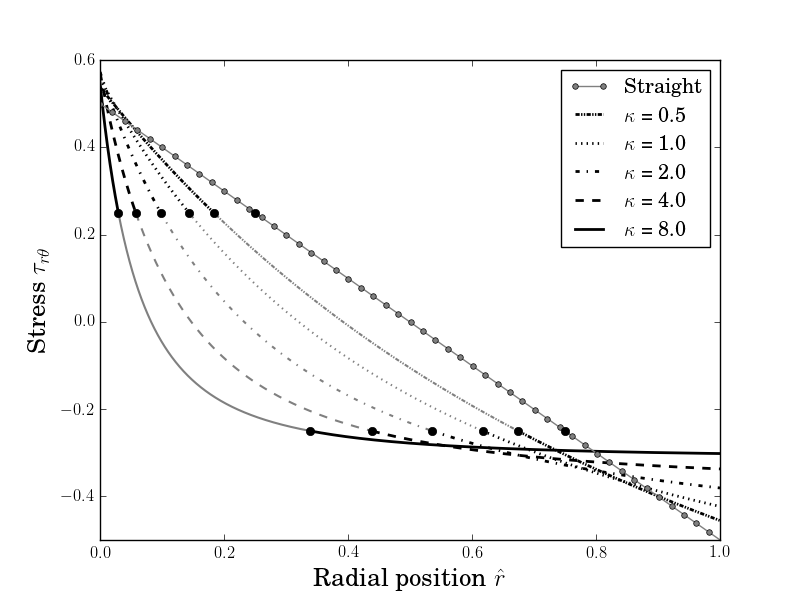} }}%
\caption{
Stress profiles as a function of radial position, with the position of the yield surfaces, at which the magnitude of the stress  is equal to the Bingham number $B$, marked with dots. 
Thinner curves in the plug regions signify that the stress is not formally defined there.
(a) For fixed channel curvature $\kappa = 2$ (for which $B_c = 0.8$) with different Bingham numbers $B$.
(b) For fixed Bingham number $B=0.5$, with different values of the channel curvature $\kappa$. 
 }%
\label{fig:stress}%
\end{figure}

Equation~(\ref{eq:stress}) shows that the shear stress decreases from the inner to the outer wall, since the pressure gradient is greatest at the inner wall. 
Figure \ref{fig:stress}(a) shows the profile of stress in a channel with $\kappa = 2$ for different values of the Bingham number $B$.
As $B$ increases towards $B_c$, the stress decreases everywhere. 

The effect on the stress of varying the channel curvature $\kappa$ is significant (Fig.~\ref{fig:stress}(b)). As $\kappa$ decreases, the plug width increases and the stress profile becomes straighter, approaching the linear profile found in a straight channel. As $\kappa$ increases the stress on the inner wall increases slightly and decreases significantly on the outer wall, resulting in a smaller region of unyielded fluid, in agreement with Fig. \ref{fig:pluglength}.

\subsection{Flux}

\begin{figure}
\centering
\includegraphics[width=7.5cm]{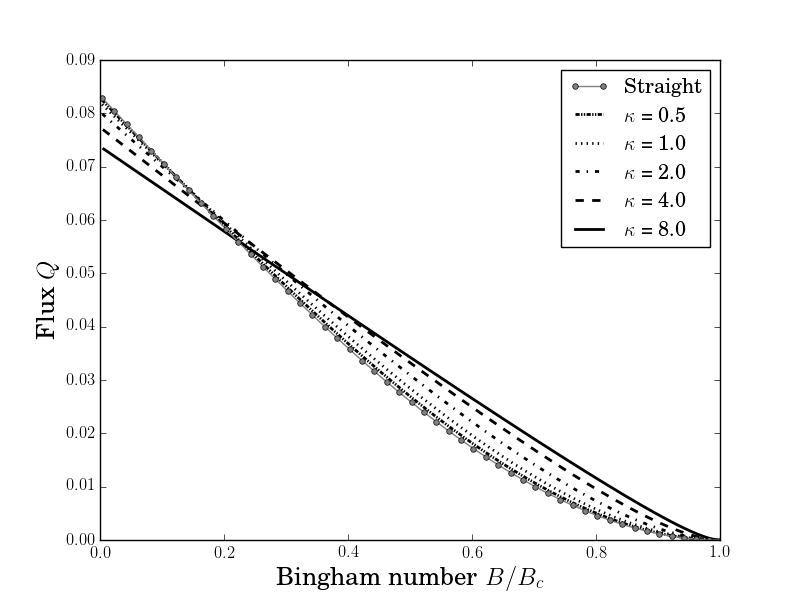}
\caption{The flux through the channel as a function of Bingham number $B/B_c$ for different channel curvatures $\kappa$.
}
\label{fig:flux}%
\end{figure}

A useful quantity to predict is the amount of fluid that flows through the channel. We calculate the one-dimensional flux $Q$, i.e. the amount of fluid which crosses a particular cross-section per unit time, by integrating the velocity profile (eq.~(\ref{eq:velprof})) with respect to radial position:

\begin{equation}
\begin{split}
Q &= \int_{\frac{1}{\kappa}}^{\frac{1}{\kappa} + 1} u_{\theta} {\rm d}r  \\
  &= \int_{\frac{1}{\kappa}}^{r_i} u_{\theta} {\rm d}r  +  \int_{r_i}^{r_o} u_{\theta} {\rm d}r +  \int_{r_o}^{\frac{1}{\kappa} + 1} u_{\theta} {\rm d}r  \\
&= B^{+} \left( \frac{{r_i}^2}{2}\left( \frac{1}{2}-\ln\left( r_i \kappa \right) \right) 
              - \frac{{r_o}^2}{2}\left( \frac{1}{2}+\ln\left( r_i \kappa \right) \right) -\frac{1}{4 \kappa^2} + \frac{r_o^2 r_i^2 \kappa^2}{4}\right) \\
&+B^{-} \left( r_o^2 \ln\left(\frac{{r_o}}{1/\kappa + 1}\right) + \frac{(1/\kappa + 1)^2}{4} - \frac{{r_o}^4}{4 (1/ \kappa + 1)^2} \right).
\end{split}
\end{equation}
Recognising that the flux is the area beneath the velocity curves in Fig.~\ref{fig:velocity}, we  expect $Q$ to tend to zero as the Bingham number approaches its critical value $B_c$, while for $B = 0$ the flux $Q$ is the value for a Newtonian fluid in the same channel. Fig.~\ref{fig:flux} shows that this is indeed the case.

Fig.~\ref{fig:flux} also shows that at low Bingham number the flux is greatest for weakly-curved channels: increasing curvature of the channel reduces the amount of material moving through the channel.
However, just as for the maximum velocity (Fig. \ref{fig:maxvelpos}(b)), at a moderate value of $B/B_c$ close to $0.2$ there is a crossover, and the flow through a curved channel is greater for given $B/B_c$. (Note that $B_c$ depends on the channel geometry so this is not equivalent to an increase of flux due to increased channel curvature for fixed $B$.) As the curvature increases the flux becomes  almost linear in $B/B_c$.

\section{Flow from a straight to a curved channel}
\label{sec:simdat}

\begin{figure}[ht]
\centerline{
\includegraphics[width=0.7\textwidth]{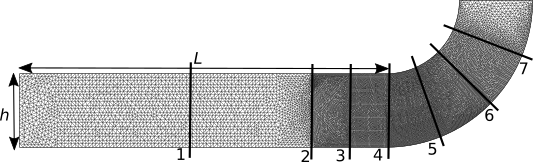}
}
\caption{The channel geometry and finite element mesh used in the simulations to examine the transition between the velocity profile in a straight and a curved channel.
The cross-sections 1-7 are used to probe the development of the velocity profile as fluid flows through the channel. The mesh in the figure is made of around $30,000$ triangles, approximately 5 times coarser than the  meshes used in the simulations. 
}
\label{fig:simsetup}
\end{figure}

We have derived the velocity profile for a Bingham fluid in a curved channel (eq.~(\ref{eq:velprof})), but the question remains as to how this profile is established when fluid enters such a channel. 
We therefore consider a geometry in which a straight channel is connected to a curved channel of the same width (Fig. \ref{fig:simsetup}). Fluid in the straight section, far from the join, flows with the usual profile (eq. \ref{eq:sc_vel}) with yield surfaces at $y_c = 1/2 \pm B/4$, while fluid in the annular section, again far from the join, flows with the velocity profile in eq. (\ref{eq:velprof}). In between, there is a transition region whose length may depend on Bingham number $B$ and/or channel curvature $\kappa$. 
The flow is steady, but nonetheless we require a numerical solution of the governing equations, described below, to determine the flow in the transition region. 


\subsection{Simulation method}

Simulations were performed with the finite element software, FreeFem++ \cite{hecht2012new}. The Stokes equations are written using the weak formulation \cite{aposporidis2011mixed} with velocities in finite element space $P3$ and pressures in $P1$. 

We model a Bingham fluid as a generalised Newtonian fluid with viscosity given by the Papanastasiou approximation \cite{papanastasiou1987}, in which an exponential function is used to smooth over the singularity in viscosity at the yield surfaces:
\begin{equation}
\eta = \left( 1 + \frac{B}{|\dot{\gamma}|}(1- \exp(-m|\dot{\gamma}|)) \right).
\label{eq:viscosity}
\end{equation}
Here $m$ is a regularisation parameter and $|\dot{\gamma}| = \sqrt{(\dot{\gamma}:\dot{\gamma})/2}$ denotes the second invariant of the rate-of-strain tensor. 

We first determine an appropriate value of $m$ by simulating flow in a curved channel with width $h=1$, where eq.~(\ref{eq:velprof}) applies. For $m = 5000$ the sum of the errors in the velocities is less than $10^{-5}$. The simulation commences with $B=0$, and then $B$ is slowly increased over 100 iterations, allowing a profile of viscosity to develop according to eq.~(\ref{eq:viscosity}). 
 


We choose the straight channel length to be $L = 5$, i.e. five channel widths long. The mesh for the joined straight and curved channel consists of $154,392$ triangles, with highest density close to the walls of the channel and around the region where the channels meet (Fig.~\ref{fig:simsetup}). We take a unit pressure gradient, which is achieved by setting the inflow pressure to $p_{in} = 5 + \frac{\pi}{2} \left(\frac{1}{\kappa} + \frac{1}{2} \right)$ and outflow pressure to zero, and record two measures of the flow to determine the properties of the transition region:
\begin{itemize}
\item the velocity profile across different cross-sections of the channel. From this we can find the distances upstream and downstream of the join between channel sections at which the velocity profiles coincide with the analytic ones.
\item the area of the unyielded plug region, normalized by the channel area. This gives a broader indication of the disruption to the flow caused by the transition to a curved channel.
\end{itemize}

The cross-sections are taken at three positions along the straight channel (Fig.~\ref{fig:simsetup}), at distances $2.5h$, $0.5h$ and $0.25h$ upstream of the join; at one cross-section where they join; and at three further cross-sections at angles of $\pi/8$, $\pi/4$ and $3\pi/8$ from the join in the curved channel.




\begin{figure}
\centering
\subfloat[]{{\includegraphics[width=7.5cm]{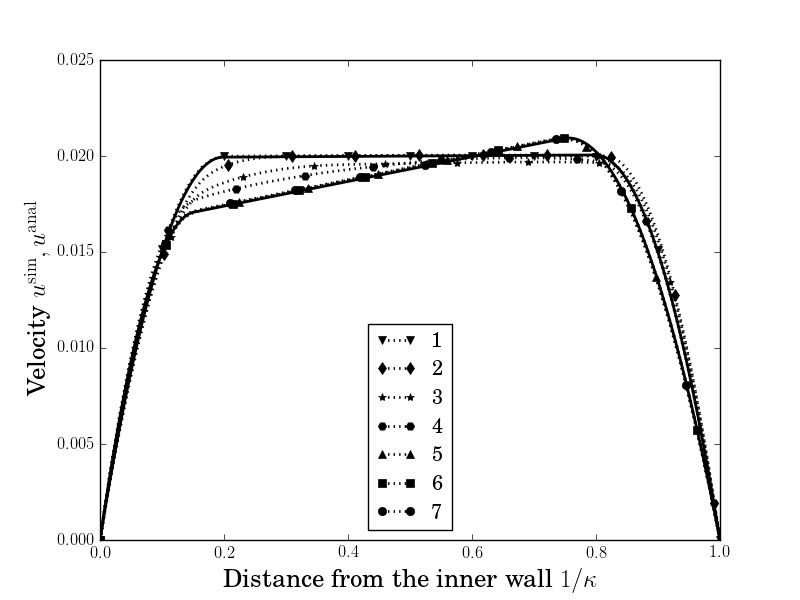} }}%
\qquad
\subfloat[]{{\includegraphics[width=7.5cm]{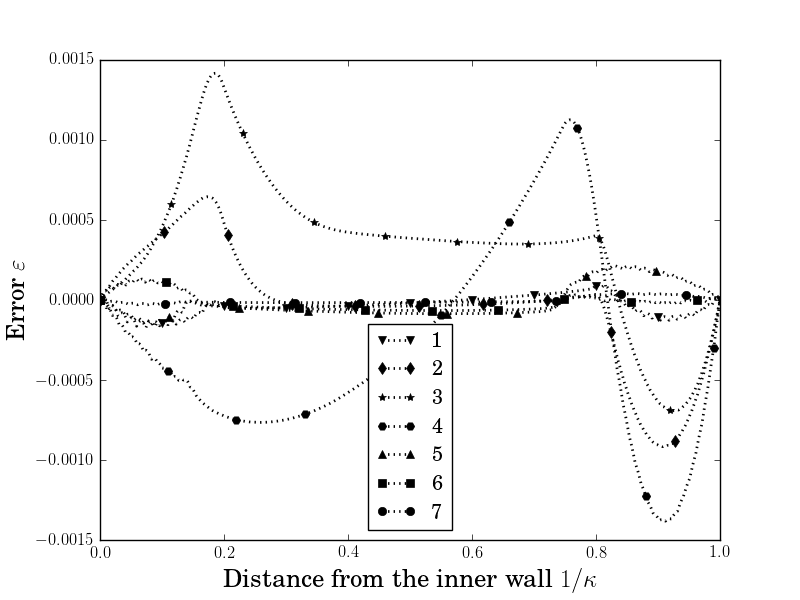} }}%
\caption{ (a) Velocity profiles in the transition region, shown at the numbered cross-sections indicated in Fig. \ref{fig:simsetup}. Channel curvature is $\kappa = 0.4$ and Bingham number is $B= 0.6$.
(b) Difference between the simulation data and the appropriate analytical velocity profile.
}%
\label{fig:vel_transition}%
\end{figure}

\subsection{Velocity profiles}

We first set the Bingham number to $B = 0.6$ and show the velocity profiles for fixed channel curvature $\kappa = 0.4$ in Fig.~\ref{fig:vel_transition}(a). 
At cross-section $1$ the velocity profile takes the form of the straight channel velocity profile,
symmetric about the channel centerline.
At cross-sections $2$ and 3, closer to the start of the curved region, there is a clear deviation from this profile and the beginning of a smooth transition towards the curved channel velocity profile, with fluid moving more slowly in the inner yielded region. By cross-section $5$ the velocity profile almost overlaps the curved channel velocity profile,  with just a small discrepancy near the outer yield surface. In cross-sections $6$ \& $7$ it isn't possible to see a difference between the curves.

A more precise indication of convergence is given by the discrepancy in the velocity along each cross-section, defined as
\begin{equation} \label{eq:error}
\varepsilon = u^{\rm sim} - u^{\rm anal},
\end{equation}
where the superscript $\,^{\rm anal}$ refers to the straight channel profile for cross-sections 1 to 3 and to the curved channel profile for cross-sections 4-7.
Fig. \ref{fig:vel_transition}(b) shows 
that the main differences occur in the yielded regions,  close to the yield surfaces, and particularly (but not unexpectedly) around the join (cross-sections 3 and 4) between the two channels.  Cross sections 1 \& 5-7 show very small values for  $\varepsilon$, indicating that a distance $2.5h$ or an angle $\pi/8$ away from the join the fluid is moving with an unchanging analytically-predictable profile.

\subsection{Yielded regions}

\begin{figure}
\centering
\subfloat[$B = 0.2$, $\kappa = 0.22$]{{\includegraphics[width=4cm]{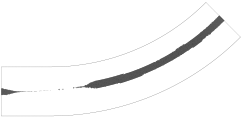} }}
\qquad
\subfloat[$B = 0.4$, $\kappa = 0.22$]{{\includegraphics[width=4cm]{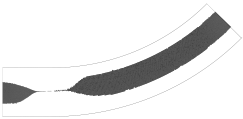} }}
\qquad
\subfloat[$B = 0.6$, $\kappa = 0.22$]{{\includegraphics[width=4cm]{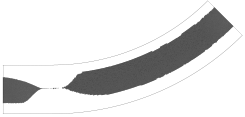} }}
\qquad
\subfloat[$B = 0.2$, $\kappa = 0.40$]{{\includegraphics[width=4cm]{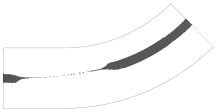} }}
\qquad
\subfloat[$B = 0.4$, $\kappa = 0.40$]{{\includegraphics[width=4cm]{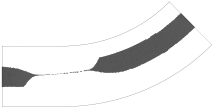} }}
\qquad
\subfloat[$B = 0.6$, $\kappa = 0.40$]{{\includegraphics[width=4cm]{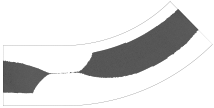} }}
\qquad
\subfloat[$B = 0.2$, $\kappa = 0.66$]{{\includegraphics[width=4cm]{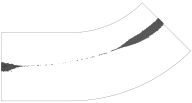} }}
\qquad
\subfloat[$B = 0.4$, $\kappa = 0.66$]{{\includegraphics[width=4cm]{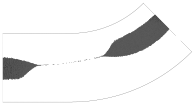} }}
\qquad
\subfloat[$B = 0.6$, $\kappa = 0.66$]{{\includegraphics[width=4cm]{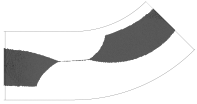} }}
\caption{ The outline of the plug region between cross sections 2 and 6 is shown for three different values of the Bingham number $B$ and three different values of the channel curvature $\kappa$ . Note how the fluid yields just downstream of where the straight channel meets the curved channel.
} 
\label{fig:sim_data}%
\end{figure}


Fig. \ref{fig:sim_data} shows examples of the shape of the unyielded regions as the fluid moves from a straight to a curved channel for different Bingham numbers $B$ and curvatures $\kappa$. Just after leaving the straight part of the channel the plug region narrows until  the fluid is yielded. The plug then reforms over roughly the same distance in the curved part of the channel. The distance over which the fluid yields increases as the annulus curvature $\kappa$ increases and as the Bingham number decreases. The width of the plug is smaller in the curved channel as expected from Fig. \ref{fig:pluglength}. 

\begin{figure}
\centering
\includegraphics[width=7.5cm]{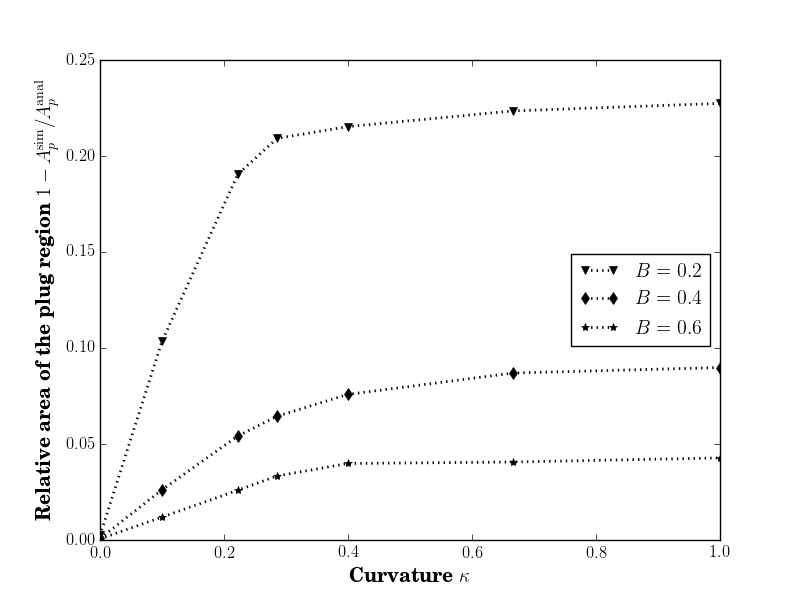} 
\caption{
The relative area of the plug region $1-A_{p}^{\rm sim}/A_{p}^{\rm anal}$ as a function of channel curvature $\kappa$ in the region between cross-sections $2$ and $6$.
The data for $\kappa = 0.0$ is from a simulation of an equivalent area of a straight channel with $h = 1.0$.
 }
\label{fig:plug_area}%
\end{figure}

If there was an abrupt change from a uniform flow in the straight part of the channel to a uniform flow in the curved part of the channel, the area of the plug region
between cross sections $2$ and $6$ would be 
\begin{equation} \label{eq:plugarea}
A_{p}^{\rm anal} = 0.5 h B \; + \;  \frac{1}{2}(r_o^2 - r_i^2)  \frac{\pi}{8}.
\end{equation} 
Clearly the actual area of the plug in this transition region is much less than this. We therefore compare the area of the plug region found in the simulations between these cross-sections,  $A_{p}^{sim}$,  with this idealised value. In the limit $\kappa \rightarrow 0$ we expect $A_{p}^{\rm sim} \rightarrow A_{p}^{\rm anal}$.


Fig. \ref{fig:plug_area} shows the relative plug area for three different Bingham numbers as a function of channel curvature. We choose to show $1-A_{p}^{\rm sim}/A_{p}^{\rm anal}$, which is a measure of the ``missing" area of unyielded fluid between cross-sections 2 and 6. Relative to our naive prediction $A_{p}^{\rm anal}$, this extra area of yielded fluid increases as the channel becomes more curved, indicating that the transition region becomes longer, in agreement with Fig. \ref{fig:sim_data}. This effect is stronger for small Bingham numbers, for which the plug is narrowest.

\section{Discussion \& Conclusion}
\label{sec:concs}

Poiseuille flows of yield stress fluids in curved channels have attracted relatively little attention. It is clear that the scenario we describe, of a pressure-driven flow in a curved channel, is difficult to implement in an experiment in isolation. Instead, it could be thought of as one element of, for example, a network of pipes conveying some yield stress fluid, which due to certain constraints must be made to turn a corner.

Our predictions allow  the effect of such a situation to be determined, for example the drop in flux associated with such a bend, as a function of the material parameters of the fluid. This work also provides a more stringent test against which to validate simulation codes for rheological models in non-trivial geometries and as a base flow which is perturbed when the flow-rate increases and secondary flows may arise.

Our main result, eq. (\ref{eq:velprof}), provides detailed insight into the dependence of the flow on the dimensions of the channel. It allows us to identify non-monotonicity in the flow, in particular in the region of maximum velocity (Fig.~\ref{fig:maxvelpos}(b)), stress (Fig.~\ref{fig:stress}(b)) and flux (Fig.~\ref{fig:flux}) in the channel.

In terms of the Bingham number $B$, we consider two extreme situations. For small $B$, the fluid behaves like a Newtonian fluid, with relatively large velocity and large stress on the inner wall $\frac{1}{\kappa}$. Such a material is likely to be ineffective at displacing a second fluid (in the example of varicose vein treatments, this second fluid is the blood that initially fills the vein), because it will be prone to instabilities such as viscous fingering.

In the other limit, as $B \rightarrow B_c$, the flow is dominated by the yield stress of the fluid and is relatively slow. The majority of the material moves as a large plug which almost completely spans the channel (Fig.~\ref{fig:pluglength}). In applications, it is this plug region which is essential for displacing another fluid. So a large Bingham number is required in varicose vein schlerotherapy.

Our result also indicates that the degree of curvature of the channel $\kappa$ affects the efficacy of a displacement flow. For a given Bingham number $B$, the width of the plug region is smaller for channels with greater curvature. In the varicose vein example, a vein that is manipulated in such a way to reduce its curvature should be treated more effectively. 

Our simulations of the transition in the velocity profile as fluid moves from a straight channel to a curved one indicate that in this region the fluid yields (Fig.~\ref{fig:sim_data}), albeit over a short distance. But in this yielded region there is likely to be a good deal of mixing between blood and foam during sclerotherapy, which again highlights the importance of keeping the vein as straight as possible during treatment.
 

\section*{Acknowledgements}

We acknowledge financial support from the UK Engineering and Physical Sciences Research Council (EP/N002326/1) and a PhD studentship from BTG.  We thank the (anonymous) reviewers for suggestions which improved the manuscript.

\bibliography{bingham_curved_08.bib} 
\bibliographystyle{unsrt}

\end{document}